\documentclass{article}
\usepackage[utf8]{inputenc}
\usepackage{amsmath}
\usepackage{graphicx,psfrag,epsf}
\usepackage[linesnumbered,ruled,vlined]{algorithm2e}
\usepackage{enumerate}

\usepackage{adjustbox}
\usepackage{booktabs} 

\begin{document}

\title{\bf Utilizing Win Ratio Approaches and Two-Stage Enrichment Designs for Small-Sized Clinical Trials}
\author{Jialu Wang*, Yeh-Fong Chen**,  Thomas Gwise**\\
*Department of Statistics, The George Washington University \\
  **Division of Biometrics IX, Center of Drug Evaluation and\\ Research, US Food and Drug Administration}
\date{}
\maketitle

\centerline{\textbf{Abstract}}
$\quad$

Conventional methods for analyzing composite endpoints in clinical trials often only focus on the time to the first occurrence of all events in the composite. Therefore, they have inherent limitations because the individual patients' first event can be the outcome of lesser clinical importance. To overcome this limitation, the concept of the win ratio (WR), which accounts for the relative priorities of the components and gives appropriate priority to the more clinically important event, was examined. For example, because mortality has a higher priority than hospitalization, it is reasonable to give a higher priority when obtaining the WR. In this paper, we evaluate three innovative WR methods (stratified matched, stratified unmatched, and unstratified unmatched) for two and multiple components under binary and survival composite endpoints. We compare these methods to traditional ones, including the Cox regression, O'Brien's rank-sum-type test, and the contingency table for controlling study Type I error rate. We also incorporate these approaches into two-stage enrichment designs with the possibility of sample size adaptations to gain efficiency for rare disease studies.

\noindent%
{\it Keywords:}  adaptive clinical trial, composite endpoints, enrichment strategy, win ratio method
\vfill

\newpage

\section{Introduction}
\label{sec:intro}

In the United States, according to the ``Rare Diseases Act of 2002" (\cite{Injection}; \cite{CDER}), there are more than 6,000 rare diseases. A rare disease is defined as a condition that affects fewer than 200,000 individuals, or 1 in 1,500 people. The development of efficient approaches to utilizing individual patient data (e.g., improved study designs and sound statistical methods) is instrumental in bringing breakthrough therapies to the market early for treating rare diseases. 
Examples of rare diseases include Gaucher disease and Neuronal ceroid lipofuscinosis 2, with the FDA recommending the use of innovative designs, including umbrella designs and single-arm historical controlled designs. In the nonmalignant hematology disease area, there are also many rare disease clinical trials, (for example, WHIM syndrome and immune thrombocytopenia) that require the careful identification of endpoints to assess the efficacy of drugs. In addition, it is not possible with many diseases to conduct well-controlled, adequately powered clinical trials for pediatric populations because of ethical concerns.

Given the concern over lacking adequate study power in conducting small-sized clinical trials, innovative designs utilizing different types of efficacy endpoints with proper statistical analyses and study-wise type I error control need to be considered. Patients are likely to be heterogeneous in rare disease clinical trials. When conducting such trials, composite endpoints can be created  by combining multiple components, either requiring all components or a certain number of components or winning on multiple endpoints (e.g., 3 out of 5). Doing so can be beneficial (\cite{Claim}) and should be considered. Furthermore, valid statistical methods are imperative to efficiently handle these types of endpoints to increase the chances of detecting treatment effect. 

In this paper, we exame statistical methods utilizing win ratio methods (WR) (\cite{Mortality}; \cite{PocockWR}; \cite{GraphWR}; \cite{GuideWR}) based on both matched and unmatched pairs. We cover different types of endpoints (i.e., survival, binary, and continuous) in our evaluations as described in Section 2.

To demonstrate the pros and cons of the WR methods, we consider different winning criteria, and results are illustrated by comparing WR methods with those via O’Brien's rank-sum-type test (\cite{OBrien}) and the contingency table. Our simulation results are shown in Section 6. Besides examining the WR methods mainly applied in the single parallel design, innovative designs such as two-stage designs, including sequential parallel comparison designs (\cite{Placebo}) and sequential enriched design (\cite{SED}), are used to provide further efficiency. Our findings are shown in Section 6. Finally, our conclusions regarding the use of the WR method for small-size clinical trials and future research plans are included in Section 7.

\section{Win Ratio Methods and Notations}

For simplicity, we consider two treatment groups: one for the study drug and the other for the control, which can be a placebo. We are interested in assessing the treatment effect that can come from any component of a composite endpoint. In our evaluation, we examine the WR performance on the continuous or survival endpoint with multiple components. For example, the test hypotheses for a composite endpoint with two binary components are as follows: 
$H_0:  p_{j,t}=p_{j,p}$ for $\forall j=1,2$, and $H_1: p_{1,t}\neq p_{1,p}$ or $p_{2,t}\neq p_{2,p}$. Similarly, the test hypotheses for a composite endpoint with three continuous components are as follows: $H_0:  E_{p,j}=E_{t,j}$ for $\forall j=1,2,3$, i.e., $\beta_{pj}=\beta_{tj}$ for $\forall j=1,2,3$, and $H_1: E_{p,1}\neq E_{t,1}$ or $E_{p,2}\neq E_{t,2}$ or $E_{p,3}\neq E_{t,3}$. Later, we also take the priority of the components' importance into consideration.

\subsection{Motivation with Toy Example}

The composite endpoints have been used in many clinical trials to increase the chances of collecting more data from many domains of a disease to increase the study power. Although this idea sounds feasible and can be useful, having a clear understanding about when a composite endpoint should be considered and how to use it properly is very important. The following is a toy example to illustrate that if the composite endpoint is not constructed wisely, the results can be misleading. 

\begin{figure}
    \centering
    \includegraphics[width=0.8\textwidth]{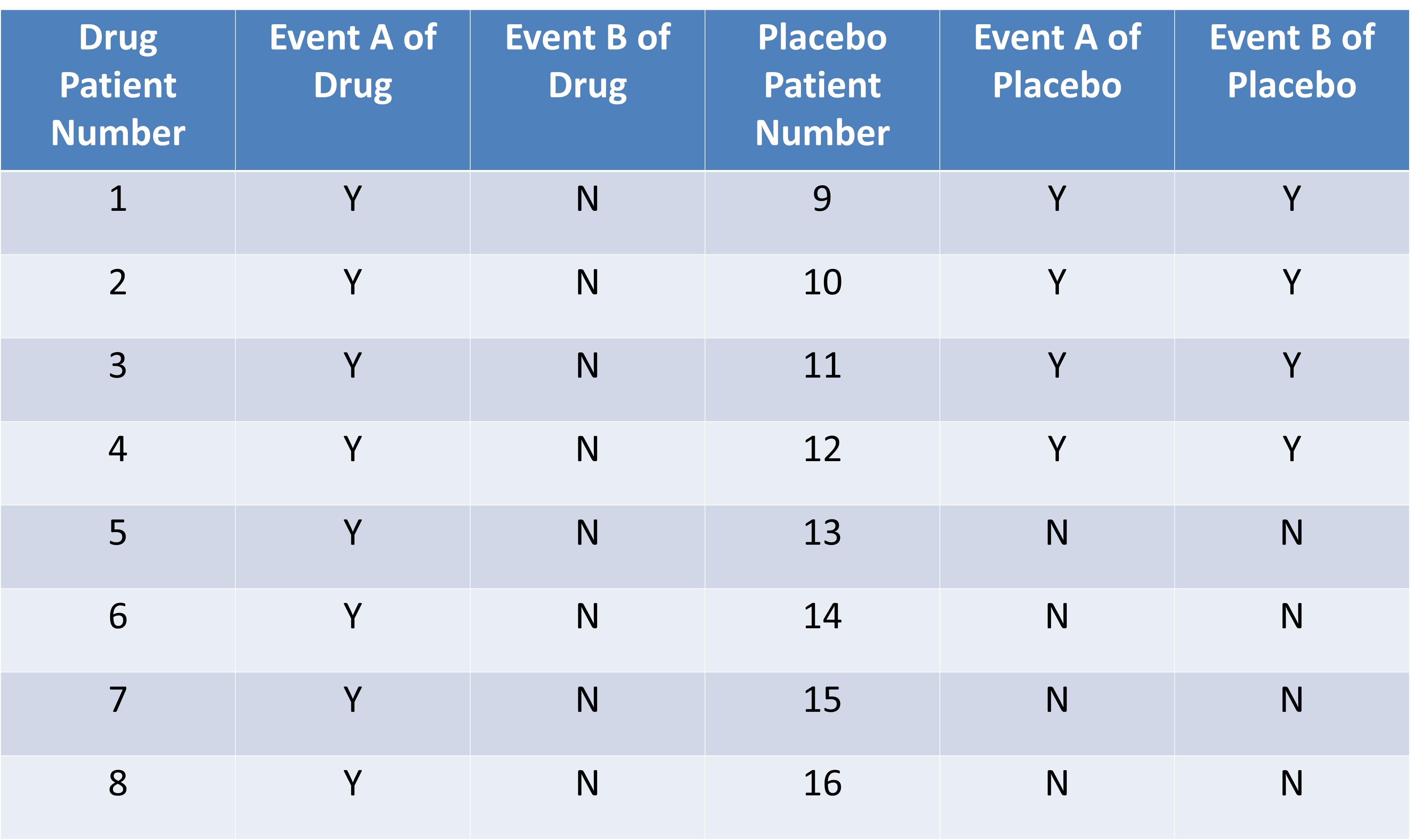}
    \caption{Toy example of composite endpoint (A or B)}
    \label{toyExample}
\end{figure}

Figure \ref{toyExample} displays our toy example with two components in a composite endpoint. 

For the drug groups, we assume that all patients respond to event A but not B. For the placebo patients, we assume half of placebo them respond to both events A and B and the other half don't respond to either event A or B. 

When we consider the composite endpoint by winning either A or B, results tell us that the drug response rate is 100$\%$ and the placebo response rate is 50$\%$. However, if we further study the two individual events, we can see that this result is mainly driven by the event A, because for the event B its drug performs worse than the placebo. In particular, it can be observed that although for Composite A or B, and also for Component A, Placebo responses 50$\%$ and Drug responses 100$\%$, for Component B Placebo responses 50$\%$ but the Drug responses 0$\%.$

In other words, if we do not consider any specific winning criteria, Event A and Event B should be equally important. Otherwise, results can be very misleading, and the study will not be powerful. 

\subsection{Literature Review for Two Types of Win Ratio methods (Unmatched and Matched)}

The idea of WRs is not new and has been extensively studied. This type of endpoint has also been utilized in many large cardiovascular and renal clinical trials (\cite{PocockWR}; \cite{NonNormalWR}). The basic idea of constructing a WR is first to pair all patients in two treatment arms and compare their performance according to pre-defined criteria to determine their winning status. At the end, combine all pairs' winning status for making the final statistical inference. These pairs can be either coming from matched (\cite{PocockWR}; \cite{WR_CI}) or unmatched samples (\cite{Mortality}; \cite{NonNormalWR}; \cite{LargeSample}; \cite{SampleSize}). More details regarding how we applied the WR methods in either unmatched or matched pairs will be discussed and illustrated in Section 3. As noted in our toy example, how all the components are prioritized in the composite endpoint will affect the performance and interpretability of the WR results. 

\section{Win Ratio Winning Criteria and Sample Size Calculation}

\subsection{Composite Endpoint With Prioritized Components}

\subsubsection{Prioritized Binary Component}
We begin the evaluation by considering the composite endpoint with two binary prioritized components. Suppose the two components we consider are death events and hospitalization. We also assume that the death event is more clinically critical than hospitalization. We theoretically derive the test statistics and confidence interval under the null hypothesis and the analytical formula for sample size calculation. 

\textbf{Notation} Let $Y_{ti}$ denote the death event for the $i$th patient who is assigned in the true treatment group (i.e., patients take the assigned drug) $T$ and assume that all patients' death events are independent. Therefore, $Y_{ti}$ follows $Bernoulli (p_t)$ identically independently (i.e., $iid.$), where $Y_{ti}=1$ represents that the $i$th patient dead and $Y_{ti}=0$ represents the patient living after the treatment. Similarly, we let $Y_{ci} $ be the indicator of the death event for $i$th patient who is assigned in the control group $C$, and $Y_{ci}$ $iid$ follows $Bernoulli (p_c)$. In addition, let $X_{ti} $ be the indicator to denote the hospitalization event for the $i$th patient under in treatment group $T$, and $X_{ti}$ $iid$ follows $Bernoulli (q_t)$. That is $X_{ti}=1$ if the $i$th patient in the treatment group requires hospitalization, and $X_{ti}=0$ if the $i$th does not. Similarly, let indicator $X_{ci} $ denote the hospitalization event for the $i$th patient under the control group $C$, and $X_{ci}$ $iid$ follows $Bernoulli (q_c)$. The principle for a comparing composite endpoint with two prioritized binary components, i.e., the winning rule of WR calculation, is specified in Figure \ref{2binaryComponent}:

\begin{figure}
    \centering
    \includegraphics[width=0.8\textwidth]{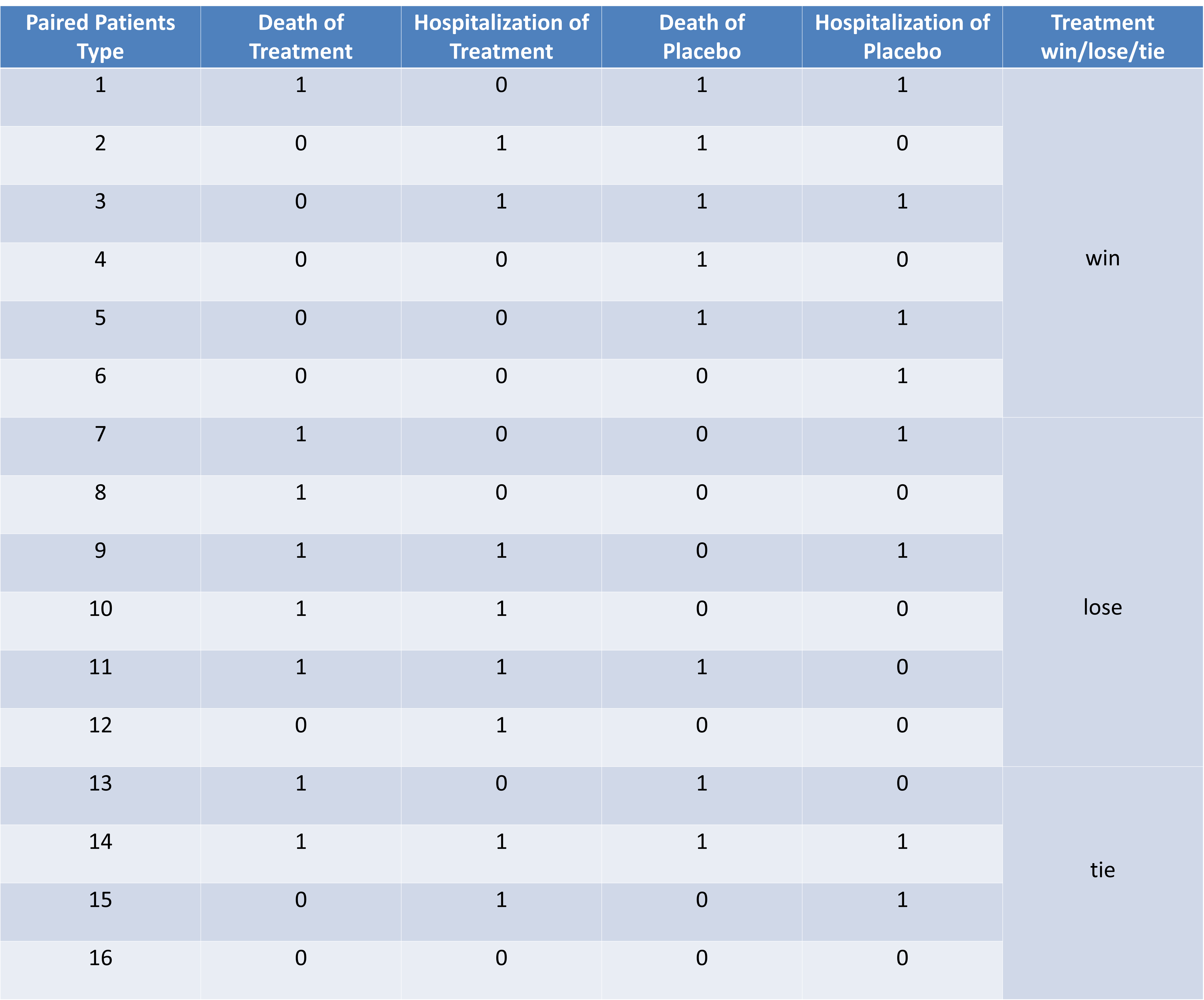}
    \caption{The comparison principle for prioritized binary endpoint with two components}
    \label{2binaryComponent}
\end{figure}

\textbf{Sample Size for Matched Win Ratio} In the previous section, we introduced the way we pair patients; either coming from matched or unmatched samples will affect the performance and interpretability of the WR results. Here we derive the asymptotic properties of WR test statistics and the sample size formula for any given Type I and power requirement. We first analyze the matched win ratio method and then the unmatched method.

First, the probability of a treatment wins under all scenarios can be derived as
\begin{align*}
p_{w}=p_t(1-q_t)p_{c}q_{c}+(1-p_t)q_{t}p_c+(1-p_t)(1-q_t)(1-(1-p_c)(1-q_c))
\end{align*}
the probability of a treatment losses under all scenarios is:
\begin{align*}
p_{l}=p_t(1-q_t)(1-p_{c})+p_{t}q_{t}(1-p_{c}q_{c})+(1-p_t)q_t(1-p_c)(1-q_c)
\end{align*}
and the probability of treatment and control ties under all scenarios is
\begin{align*}
p_{tie}=1-p_{w}-p_{l}
\end{align*}

Next, we let the binary random variable $X_i$ follow $Bernoulli(p)$, which denotes every win-loss comparison, where $X_i=1$ if treatment wins; otherwise, $X_i=0$, and 
\begin{align*}
    p=P(\text{treatment win}|\text{all non-tie pairs})
     =\frac{p_w}{1-p_{tie}}
\end{align*}

Suppose a total number of $N$ patients are randomized, and we let $n=N(1-p_{tie})$ denote the total number of non-tie units. Based on the Delta Method, we derive that
\begin{align}
    \sqrt{n}\left(\frac{\Bar{X}}{1-\Bar{X}} - \frac{p}{1-p}\right)\xrightarrow{D}N\left( 0, \frac{p^2}{(1-p)^2}\right), \quad \text{where} \quad \Bar{X}=\sum^{n}_{i=1}X_i/n.
\end{align}

It is obvious that under the null hypothesis, $p=0.5$ and $\sqrt{n}\left(\frac{\Bar{X}}{1-\Bar{X}} - 1\right)\xrightarrow{D}N\left( 0, 1\right)$. Besides, the minimum sample size $N$ required for power $\beta$ under Type I error $\alpha$ is

\begin{align}
    N=\frac{n}{1-p_{tie}} \quad \text{and} \quad 
    n= \left( \frac{Z_{\alpha}-\frac{p_a}{1-p_a}Z_{\beta}}{\frac{2p_a-1}{1-p_a}} \right)^2 
\end{align}
where $p_a$ is the proportion under alternative hypothesis.

\textbf{Sample Size for Unmatched Win Ratio}
Similar to the matched WR, we first consider all the scenarios in which treatment wins and treatment losses.

For treatment and control pair $(i, j),$ when $Y_{ti}=0, Y_{cj}=1,$ or $Y_{ti}=1, Y_{cj}=1, X_{ti}=0, X_{cj}=1$, or $Y_{ti}=0, Y_{cj}=0, X_{ti}=0, X_{cj}=1$, treatment wins. Similarly, when $Y_{ti}=1, Y_{cj}=0,$ or $Y_{ti}=1, Y_{cj}=1, X_{ti}=1, X_{cj}=0$, or $Y_{ti}=0, Y_{cj}=0, X_{ti}=1, X_{cj}=0$, control wins. 

Therefore, we can derive the test statistics for win ratio $g(\textbf{X})$ by dividing the total number of treatment wins by the total number of control wins, where $\mathbf{X}=(\overline{Y_t}, \overline{X_t}, \overline{XY_t}, \overline{Yc}, \overline{Xc}, \overline{XY_c})$ and $\overline{Y_t}=\sum^{n_1}_{i=1}Y_{ti}$, $\overline{X_t}=\sum^{n_1}_{i=1}X_{ti}$, $\overline{XY_t}=\sum^{n_1}_{i=1}X_{ti}Y_{ti}$, $\overline{Yc}=\sum^{n_0}_{j=1}Y_{cj}$, $\overline{Xc}=\sum^{n_0}_{j=1}X_{cj}$, $\overline{XY_c}=\sum^{n_0}_{j=1}X_{cj}Y_{cj}$. The $n_1$ is the total number of patients that are assigned to the treatment group, and $n_0$ is the total number of patients that are assigned to the control group, and $n_t=n_1+n_0$.

Then by the Delta Method, we can derive 

\begin{align}
\sqrt{n_t}\left(g(\mathbf{X})-g(\boldsymbol{\theta})\right) \xrightarrow{D}N\left( 0, \left(\frac{d}{d\boldsymbol{\theta}}g(\boldsymbol{\theta})\right)^{T}COV(\mathbf{X})\left(\frac{d}{d\boldsymbol{\theta}}g(\boldsymbol{\theta})\right)
\right)
\end{align}
where $\boldsymbol{\theta}=(p_t, q_t, p_{t}q_t, p_c, q_c, p_{c}q_c)$, $g(\boldsymbol{\theta})=g(E(\mathbf{X}))$ and $g(.)$ can be derived and is specified in the Appendix.

Therefore, under the null hypothesis
\begin{align}
\sqrt{n_t}\left(g(\mathbf{X})-1\right) \xrightarrow{D}N\left( 0, C_{0}^2\right),\quad
C_{0}^2=\left(\frac{d}{d\boldsymbol{\theta}}g(\boldsymbol{\theta})\right)^{T}COV(\mathbf{X})\left(\frac{d}{d\boldsymbol{\theta}}g(\boldsymbol{\theta})\right)_{| \boldsymbol{\theta}=\boldsymbol{\theta_0}}
\end{align}
where $\boldsymbol{\theta_0}=(0.5, 0.5, 0.25, 0.5, 0.5, 0.25)$.

Similarly, under the alternative hypothesis
\begin{align}
\sqrt{n_t}\left(g(\mathbf{X})-g(\boldsymbol{\theta_1})\right) & \xrightarrow{D}N\left( 0, C_{1}^2\right),
C_{1}^2=\left(\frac{d}{d\boldsymbol{\theta}}g(\boldsymbol{\theta})\right)^{T}COV(\mathbf{X})\left(\frac{d}{d\boldsymbol{\theta}}g(\boldsymbol{\theta})\right)_{| \boldsymbol{\theta}=\boldsymbol{\theta_1}}
\end{align}

where $\boldsymbol{\theta_1}=(p_{t1}, q_{t1}, p_{t1}q_{t1}, p_{c1}, q_{c1}, p_{c1}q_{c1})$.

Therefore, the minimum sample size required for power $\beta$ under Type I error $\alpha$ is
\begin{align}
    n_t=
    \left( \frac{C_0Z_{\alpha}-C_1Z_{\beta}}{g(\boldsymbol{\theta_1})-1} \right)^2. 
\end{align}

\subsubsection{Prioritized Survival Component}
In this section, we show the winning rules of matched and unmatched methods for the composite endpoint of two prioritized survival components. To further explore the pros and cons of the WR methods, traditional Cox regression in survival analysis and O’Brien's rank-sum-type test (\cite{OBrien}) are considered and incorporated. Point estimation and its corresponding confidence interval and power comparison are extensively explored via numerical studies in section \ref{numerical survival}.

\textbf{Matched Win Ratio}
We stratify patients into different strata based on their baseline covariates, and then form matched pairs on the study drug and the control. For each matched pair, according to the following criteria, we then compare each patient in the study drug group with the one matched in the placebo group is a winner or a loser and its asymptotic properties via Algorithm \ref{algo} (\cite{PocockWR}). We also note that \cite{WR_CI} proposed a closed-form variance estimator and approximate $1-\alpha$ confidence interval, which could be utilized for testing the null hypothesis.

\begin{algorithm}[]
\label{algo}
\caption{Matched Winning Rule}
\DontPrintSemicolon
  \If{stratified}{patients are stratified into $k$ different strata based on their covariates, form matched pairs within each stratum for the new treatment and the control; all pairs are then collected.}
   \Else{Form matched pairs based on the whole sample.}
  \For{every matched pair}{
  \If{ one of the two patients die}{
    \If{patient in the treatment group dies first}{Control wins (Treatment loses)}
    \If{patient in the control group dies first}{Treatment wins (Control loses)}
  }
  \Else 
  { \If{patient in the treatment group has hospitalization first}{Control wins}
    \If{patient in the control group has hospitalization first,}{Treatment wins}
    \Else{ Tie}
  }
    }
\end{algorithm}

We then obtain the following:
    \begin{enumerate}
      \item The number of patients that fall into categories: (a) new treatment patient has death first $N_a$ (b) control patient has death first $N_b$ (c) new treatment patient has hospitalization first $N_c$ (d) control patient has hospitalization first $N_d$. 
      \item $N_w=N_{b}+N_{d}$, the number of "winners" for the new treatment. $N_{\mathrm{L}}=N_{\mathrm{a}}+N_{\mathrm{c}}$, the number of "losers" for the new treatment. 
       \item The proportion $p_W$: $p_W=\frac{N_{\mathrm{w}}}{N_{\mathrm{w}}+N_{\mathrm{L}}},$ $
            p_L, p_U= p_{w}\pm1.96\left[\frac{p_{w}\left(1-p_{w}\right)}{\left(N_{w}+N_{L}\right)}\right]^{1/2}$
       \item The "WR"
            $R_W=\frac{N_{\mathrm{w}}}{N_{\mathrm{L}}}=\frac{p_w}{1-p_w},$ $
            CI_{R_W, 0.95}=\left(\frac{p_L}{1-p_L}, \frac{p_U}{1-p_U} \right)$
        \item The test statistics via a standardized normal assumption, for a significance hypothesis testing:
            $z=\left(p_{w}-0.5\right) /\left[p_{w}\left(1-p_{w}\right) /\left(N_{w}+N_{L}\right)\right]^{\frac{1}{2}} \label{z significance hypothesis}$
    \end{enumerate}

\textbf{Unmatched Win Ratio} From \cite{Mortality} and \cite{PocockWR}, we use the stratified Finkelstein and Schoenfeld (FS) test and derive the corresponding power by simulations. It proceeds as follows
	\begin{enumerate}
        \item Stratify patients into $k$ strata and let $A_k$ denote $n_k$ patients in the $kth$ strata.
        \item Irrespective of treatment group, compare all possible pairs of patients $i$, $j$ to determine whether patient $i$ is a winner, loser, or tie. 
        \item Calculate $N_w$ and $N_L$ via the same way as in the matched method.
        \item Define $u_{ij}$ and assign $u_{ij}= +1, -1, \text{0}$ according to winning status of patient $i$ (i.e., winner, loser, or tie).
        \item Within each stratum, calculate $U_{i}$ where for $i \in A_{k}, U_{i}=\sum_{j \in A_{k}} u_{i j}$. It will be a positive integer if patient $i$ wins more often than losses compared with all other patients.
	\end{enumerate}

We calculate the WR $R_w$ and test statistics $z$ as follows:
$$R_w=N_{\mathrm{w}}/N_{\mathrm{L}}, \quad
z=T / V^{1 / 2}, \quad T=\sum_{k}\sum_{i \in A_{k}} D_{i} U_{i},\quad
            V=\sum_{k}\frac{m_{k}(n_{k}-m_{k})}{n_{k}n_{k}-1}\sum_{i \in A_{k}} U_{i}^{2}$$
where $D_i=1$ for subjects in the new group and $D_i=0$ for patients in the standard group.

For hypothesis testing, we also utilize the standardized normal statistics $z$ in the equation \ref{z significance hypothesis} of Algorithm 1. For the confidence interval (CI) and power, we first calculate $lnR_w$ and its approximate standard error $s=lnR_w/z$. Then we have $CI_{lnR_{w}, 0.95}=\left(lnR_{w,L}, lnR_{w,U}\right)=\left(lnR_w-1.96s, \ lnR_w+1.96s\right)$, and thus $CI_{R_w, 0.95}=(e^{lnR_{w,L}}, \ e^{lnR_{w,U}}).$

For the unstratified unmatched WR, we follow the same step as the stratified unmatched WR method except for the stratification. We call it the unstratified unmatched WR method.

\textbf{Cox Regression} In this section, we use Cox regression to analyze the time to the first event of the composite endpoint. For example, in a typical Cox regression equation 
	\begin{align}
	    h(t)=h_{0}(t)exp(\beta_{t}x_{t}+\beta_cx_{cov})
	\end{align}
The $h(t)$ is hazard rate at given time $t$, where $t=min(E_{d}, E_{hos})$.
The $x_t$ is an indicator representing whether the patient is in the treatment group, and $x_{cov}$ are patients' baseline covariates. $h_{0}(t)$ is the baseline hazard, which does not depend on treatment indicator $x_t$ and covariates $x_{cov1}, x_{cov2}$. Finally, $\beta_{t}$ is the expected log hazard ratio (HR) that compares the risk of a patient in treatment to those in the control arm for both death and hospitalization events, and we are interested in testing whether $\beta_{t}$ is $0$ or not under required Type I error.

\textbf{O'Brien's Rank-Sum-Type Test}
Peter C. O'Brien proposed a rank-sum-type test in ``Procedures for Comparing Samples with Multiple Endpoints (1984)," and we incorporate it within the context of composite endpoint as follows:  
	\begin{enumerate}
	    \item Let $Y_{ijk}$ represent the $k$th variable for the $j$th subject in Group $i$, where $(k = 1, ..., K, j= 1, ... , n_i, i = 1, ..., I)$. $Y_{ijk}$ is defined so that large values are better than small values for each $k = 1 ..., K.$ (e.g., $k$ is death or hospitalization, $j=1,...,n$, $i$ is treatment or control group.)
        \item Let $R_{ijk}$ represent the rank of $Y_{ijk}$ among all values of variable $k$ in the pooled set of $I$ samples. Define $S_{ij}$ as the sum of the ranks assigned to the $j$th person in sample $i$. 
        \item Perform a one-way Analysis of Variance (ANOVA) on the $S_{ij}$ values. 
	\end{enumerate}

\subsection{Component Endpoint With Equally Important Continuous Components}
To generalize the use of the WR method in a composite endpoint with more than two components, we consider the situations in which a composite endpoint has multiple equally important components. For example, a composite endpoint with three equally important continuous components has notations described as follows

Suppose $y_{p, j, i}$ is the $i{th}$ patient's time to its $jth$ component improvement in the placebo group, $y_{t, j, i}$ is the $i{th}$ patient's time to its $jth$ component improvement in the treatment group, and $y_{base}$ is a baseline. We identify the indicators of successful improvement for patients in the placebo group via the following indicators:
\begin{align*}
    \mathcal{I}_{p,j,i}=
    \left\{
  \begin{array}{rcr}
    1& \quad y_{p,j,i}/y_{base,i}<c_{t} \\
    0& \quad y_{p,j.i}/y_{base,i}\geq c_{t}\\
  \end{array}
\right.
\quad
    \mathcal{I}_{p,i}=
    \left\{
  \begin{array}{rcr}
    1& \quad \sum_{j=1}^{3}\mathcal{I}_{p,j,i} \geq 1 \\
    0& \quad \sum_{j=1}^{3}\mathcal{I}_{p,j,i} =0, \\
  \end{array}
\right.
\end{align*}
where $\mathcal{I}_{p,j,i}$ is an indicator that implies whether the $ith$ patient in placebo group successfully improves on the $j^{th}$ component with cutoff $c_{t}$, and
$\mathcal{I}_{p,i}$ is an indicator that implies whether the $ith$ patient in placebo group successfully improves on at least one component. Similarly, we identify the indicators of successful improvement $\mathcal{I}_{t,j,i}$ and $\mathcal{I}_{t,i}$ for patients in the treatment group via the following indicators:
 \begin{align*}
    \mathcal{I}_{t,j,i}=
    \left\{
  \begin{array}{rcr}
    1& \quad y_{t,j,i}/y_{base,i}<c_{t} \\
    0& \quad y_{t,j,i}/y_{base,i}\geq c_{t}\\
  \end{array}
\right.
\quad
    \mathcal{I}_{t,i}=
    \left\{
  \begin{array}{rcr}
    1& \quad \sum_{j=1}^{3}\mathcal{I}_{t,j,i} \geq 1 \\
    0& \quad \sum_{j=1}^{3}\mathcal{I}_{t,j,i} =0 .\\
  \end{array}
\right.
\end{align*}

\textbf{Matched Win Ratio} The logic here is similar to the Algorithm \ref{algo} except for some modification, especially the way to define the winner in every matched pair comparison. We stratified patients into different strata based on their baseline covariates and then form matched pairs on the study drug and the control. For each matched pair, we determine that the patient in the study drug is a winner or a loser by the following rule
	    \begin{enumerate}
	        \item Calculate the total number of successful improvements for each patient in placebo, i.e., calculate $\sum_{j=1}^{3}\mathcal{I}_{p,j,i}$, $i=1,.., n_{0}.$
	        \item Calculate the total number of successful improvements for each patient in treatment, i.e., calculate $\sum_{j=1}^{3}\mathcal{I}_{t,j,i}$, $i=1,.., n_{1}.$
	        \item Within each pair, if the total number of successful improvements for the patient in treatment is greater than that for the patient in placebo, treatment wins. 
	        \item Within each pair, if the total number of successful improvements for the patient in treatment is less than that for the patient in placebo, control wins.
	        \item Otherwise, tie. 
	    \end{enumerate}
Calculate $N_{w}$, the number of winners, and $N_{L}$, the number of losers for the study drug. The test statistics is the same as the one in Algorithm \ref{algo}.

\textbf{Unmatched Win Ratio} The procedure here is the same as the unmatched WR method for the composite endpoint with the prioritized survival components. However, like the above matched WR for continuous components, the rule to define the winner in every matched pair comparison is completely different and should follow the winning rule in the new matched WR.

\textbf{Contingency Table} For evaluating the advantage of using WR methods, we construct the following traditional contingency table and perform hypothesis test via odds ratio as shown below.
\begin{table}
\centering
\begin{tabular}{cccc}
\hline
          & Success  & Failure  & Total    \\ \hline
Treatment & $n_{11}$ & $n_{10}$ & $n_{1.}$ \\
Placebo   & $n_{01}$ & $n_{00}$ & $n_{0.}$ \\
Total     & $n_{.1}$ & $n_{.0}$ & $N$      \\ \hline
\end{tabular}
\end{table}
where $n_{11}=\sum^{n_{1.}}_{i=1}\mathcal{I}_{t,i}$, $n_{10}=\sum^{n_{1.}}_{i=1}(1-\mathcal{I}_{t,i})$,
$n_{01}=\sum^{n_{0.}}_{i=1}\mathcal{I}_{p,i}$,and $n_{00}=\sum^{n_{0.}}_{i=1}(1-\mathcal{I}_{p,i})$. Therefore, the test statistic and its distribution is
$$
    \hat{OR}=\frac{n_{11}n_{00}}{n_{10}n_{01}},\quad
    log(\hat{OR}) \sim N(0, \hat{se}),\quad
    \hat{se}=\sqrt{ \frac{1}{n_{11}}+\frac{1}{n_{10}}+\frac{1}{n_{01}}+\frac{1}{n_{00}}}
$$

$\quad$

\section{Sequential Enriched Design}

\begin{figure}
    \centering
    \includegraphics[width=0.8\columnwidth]{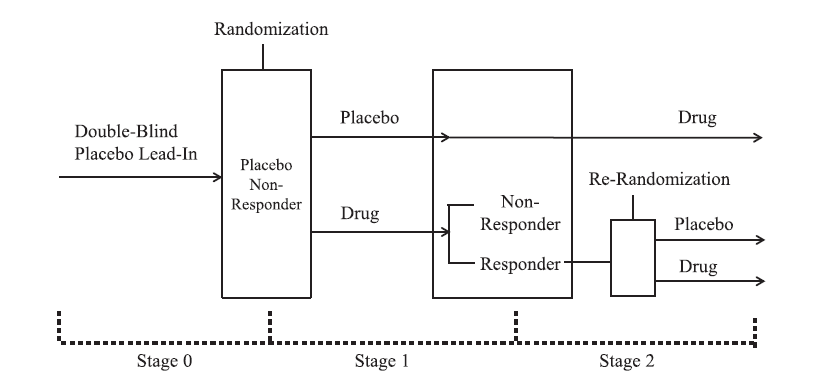}
    \caption{SED procedure}
    \label{SEDProcedure}
\end{figure}

To further enhance trial efficacy, two-stage designs can be considered for rare disease clinical trials. In our illustration, we considered sequential enriched design (SED). As seen in Figure \ref{SEDProcedure}, SED has two stages. However, before patients are randomized to the first main stage, a placebo lead-in phase is built in to determine their placebo response status. The first major stage of SED is a traditional parallel design, and at the end of the first stage, only patients in the drug group of Stage 1 and are also responders will be further rerandomized to the second stage. The goal of SED is to only study patients who are both placebo non-responders and drug responders.  

We use $c_{s0}$ to denote the cutoff for determining placebo nonresponders, i.e., if $y_{pj,i}/y_{base,i}>c_{s0}$ for $\forall j=1,2,3$, then the $ith$ patient is a placebo nonresponder. Let $c_{s1}$ be the cutoff for determining drug nonresponders, i.e., if $y_{j,i}/y_{base,i}>c_{s1}$ for all $j=1,2,3$, the $ith$ patient is drug nonresponder.

\begin{table}
\caption{Distribution of overall patient population} \label{SEDDistribution}
\begin{adjustbox}{width=0.8\columnwidth}
\begin{tabular}{@{}cccc@{}}
\toprule
Proportion                & True drug responders    &  True drug nonresponders  \\
True placebo responder    & $p_1$ & $p_2$       \\ \midrule
True placebo nonresponder & $p_3$ & $p_4$   \\
\bottomrule
\end{tabular}
\end{adjustbox}
\end{table}

As shown in Figure \ref{SEDDistribution}, the overall patient population is composed of four subpopulations according to the treatment patients receive and whether they respond to the received treatment: drug responders and placebo responder $p_1$, drug nonresponders and placebo responders $p_2$, drug responders and placebo nonresponders $p_3$, and drug nonresponders and placebo nonresponders $p_4$. Note that in SED, the target patient population is the type of patients with $p_3$ probability. 

\section{Data Generation}

\subsection{Composite Endpoint With Two Survival Components}

We used `coxed' package in R statistical software to generate survival time response. For simplicity, we illustrate our idea by only considering two components, e.g., death and hospitalization.
\\
\textbf{Time to the Component Improvements With Less Clinical Importance}
$$E_{hos}=H_{0}^{-1}[-\log (u) \exp (-X \beta_{hos})]$$
where $X=(x_{t}, x_{cov1},x_{cov2})$,  $\beta_{hos}=(\beta_{t}, \beta_{cov1}, \beta_{cov2}).$ The $x_t$ is an indicator of whether the patient is in the treatment group.
$\beta_{t}$ is the expected log HR that compares the risk of a patient in treatment to that in control for hospitalization. The drug is effective if $\beta_{t}>0$. 
$\beta_{cov1}$ and $\beta_{cov2}$ are coefficients of covariate $x_1$ and $x_2,$ respectively. $u$ is a random sample drawn from a standard uniform distribution $\mathcal{U}[0,1]$.
\\
\textbf{Time to the Component Improvements With More Clinical Importance}
    $$E_{d}=H_{0}^{-1}[-\log (u) \exp (-X \beta_d)]$$
where $X=(x_{t}, x_{dhraito}, x_{cov1},x_{cov2})$, $\beta_{d}=(\beta_{t}+\beta_{in}, \beta_{dhratio}, \beta_{cov1}, \beta_{cov2}).$ $\beta_{in}$ is expected log HR that describes the difference between the risk of a patient for death and hospitalization in treatment group. Therefore, $\beta_t^\prime =\beta_t+\beta_{in}$ is the expected log HR that compares the risk of a patient in the treatment to that in control for the death event. The $x_{dhraito}$ is a standardized random variable that describes the strength of the relationship between risk of death and hospitalization for each patient without treatment effect. The $\beta_{dhraito}$ describes the strength of the relationship between $E_d$ and $x_{dhratio}$. $\beta_{dhraito}=0$ indicates that the patient's risk of hospitalization is equal to their risk of death in the control group.

\subsection{Composite Endpoint With Three Equally Important Continuous Components and Repeated Measurements}

\textbf{Time to Patient's Three Component Improvements in the Placebo Group}
$$
    y_{base}=\beta_{cov1}x_{1}+\beta_{cov2}x_{2} \quad
    y_{pj}=\beta_{pj}(1-x_{t})+y_{base}+\epsilon_{pj} \quad j=1,2,3
$$
 The $y_{base}$ is a baseline vector and    
 $\beta_{cov1}$ and $\beta_{cov2}$ are coefficients of covariate vectors $x_1$ and $x_2$, respectively.
 In addition, $y_{pj}$ is a vector that stores the time to the $jth$ component improvement of patients who are in the placebo group and
 $x_t$ is an indicator vector that shows whether patients are in the treatment group or placebo. The
 $\beta_{pj}$ is the placebo effect that may reduce a patient's time to the $jth$ component improvement in placebo to that in baseline. It is effective if $\beta_{pj}<0$. The
 $\epsilon_{pj}$ is the randomness that corresponds to the $jth$ placebo response.
 \\
\textbf{Time to Patient's Three Component Improvements in the Treatment Group}
\begin{align*}
    y_1=\beta_{t1}x_{t}+y_{base}+\epsilon_{t1},\,
    y_2=(\beta_{t1}+\beta_{in2})x_{t}+y_{base}+\epsilon_{t2},\,
    y_3=(\beta_{t1}+\beta_{in3})x_{t}+y_{base}+\epsilon_{t3} 
\end{align*}

$\beta_{t1}$ is drug effect that reduces a patient's time to the $1^{st}$ component improvement in treatment to that in baseline. It is effective if $\beta_{t}<0$. The 
$\beta_{in2}$ describes the difference of drug efficacy between the $1^{st}$ and $2^{nd}$ components in treatment group, i.e, $\beta_{t2} =\beta_{t1}+\beta_{in2}$ is the drug effect that reduces a patient's time to the $2^{nd}$ component improvement in treatment to that in baseline. In addition,
$\beta_{in3}$ has a similar definition to $\beta_{in2},$ and
$\epsilon_{tj}$ for $j=1,2,3$ is the randomness that corresponds to the $ith$ treatment response.

\section{Numerical Study}

We have performed simulations to examine the utility of the WR in the survival endpoints and the continuous endpoints. Our simulation results are shown in Sections 6.1 and 6.2. In each simulation, we obtained WRs and test results of Type I error and study power, and our simulation results for the survival endpoint are shown in Section 6.2.  

\subsection{Survival Composite Endpoint With Two Components under Parallel Design}
\label{numerical survival}
We let $\beta_{cov1}=-0.5$, $\beta_{cov2}=0.5$, $x_{cov1}, x_{cov2} \sim$ $Bernoulli(0.5).$ $x_{dhratio} \sim$ $Uniform(0,1),$ The Table \ref{Numeric_Distribution} shows distribution of patients in the four strata we generated
\begin{table}
\caption{Distribution of patients}
\centering
\begin{adjustbox}{width=0.4\columnwidth}
\begin{tabular}{@{}ccccc@{}}
\toprule
Stratum                                                           & 1    & 2    & 3    & 4    \\ \midrule
\begin{tabular}[c]{@{}c@{}}Percentage of \\ patients\end{tabular} & 24.5 & 23.6 & 26.5 & 25.4 \\ \bottomrule
\end{tabular}
\end{adjustbox}
\label{Numeric_Distribution}
\end{table}
We estimate the HR based on Cox regression and calculate the WR for our proposed SED and analyses. In addition, we calculate the corresponding confidence intervals and Type I error as well as power via the exact methods.

\textbf{Type I error} When under the null hypothesis, a drug has no effect such that every patient is equally likely to have hospitalization/death in the treatment and control groups time. We show that either HR or win ratios are close to 1 and the Type I errors are controlled for all examined methods. Our results are displayed in Table \ref{TypeI1} and Table \ref{TypeI2}.

\begin{table}
\caption{The estimation of treatment effect of different sample sizes}
\centering
\begin{adjustbox}{width=0.9\columnwidth}
\begin{tabular}{@{}llll@{}}
\toprule
Estimation (confidence interval)                                                     & N=60                                                               & N=100                                                              & N=200                                                                                                                       \\ \midrule
\begin{tabular}[c]{@{}l@{}}HR\end{tabular}  & \begin{tabular}[c]{@{}l@{}}1.05 (0.60, 1.83)\end{tabular} & \begin{tabular}[c]{@{}l@{}}1.02 (0.67, 1.54)\end{tabular} & \begin{tabular}[c]{@{}l@{}}1.02 (0.76, 1.37)\end{tabular} \\
\begin{tabular}[c]{@{}l@{}}Stratified, matched WR\end{tabular}    & \begin{tabular}[c]{@{}l@{}}1.01 (0.44, 2.33)\end{tabular} & \begin{tabular}[c]{@{}l@{}}1.01 (0.54, 1.89)\end{tabular} & \begin{tabular}[c]{@{}l@{}}1.00 (0.65, 1.53)\end{tabular} \\  
\begin{tabular}[c]{@{}l@{}}Stratified, unmatched WR\end{tabular}    & \begin{tabular}[c]{@{}l@{}}1.05 (0.54, 2.02)\end{tabular} & \begin{tabular}[c]{@{}l@{}}1.03 (0.63, 1.68)\end{tabular} & \begin{tabular}[c]{@{}l@{}}1.00 (0.72, 1.41)\end{tabular} \\ 
\begin{tabular}[c]{@{}l@{}}Unstratified, unmatched WR\end{tabular}    & \begin{tabular}[c]{@{}l@{}}1.04 (0.56, 1.90)\end{tabular} & \begin{tabular}[c]{@{}l@{}} 1.03 (0.65, 1.64)\end{tabular} & \begin{tabular}[c]{@{}l@{}} 1.01 (0.73, 1.40)\end{tabular} \\ 
\bottomrule
\end{tabular}
\end{adjustbox}
\label{TypeI1}
\end{table}

\begin{table}
\caption{Type I error comparison with $\beta_{t}=\beta_{in}=\beta_{dhratio}=0$.}
\centering
\begin{adjustbox}{width=0.6\columnwidth}
\begin{tabular}{@{}llll@{}}
\toprule
Type I error                                                         &N=60                                                               &N=100                                                              & N=200                                                                                                                   \\ \midrule
\begin{tabular}[c]{@{}l@{}}Cox regression\end{tabular}  & \begin{tabular}[c]{@{}l@{}} 0.05\end{tabular} & \begin{tabular}[c]{@{}l@{}} 0.05\end{tabular} & \begin{tabular}[c]{@{}l@{}}0.05\end{tabular} \\
\begin{tabular}[c]{@{}l@{}}Stratified, matched WR\end{tabular}    & \begin{tabular}[c]{@{}l@{}} 0.06\end{tabular} & \begin{tabular}[c]{@{}l@{}}0.06\end{tabular} & \begin{tabular}[c]{@{}l@{}} 0.06\end{tabular} \\  
\begin{tabular}[c]{@{}l@{}}Stratified, unmatched WR\end{tabular}    & \begin{tabular}[c]{@{}l@{}}0.04\end{tabular} & \begin{tabular}[c]{@{}l@{}} 0.05\end{tabular} & \begin{tabular}[c]{@{}l@{}}0.05\end{tabular} \\
\begin{tabular}[c]{@{}l@{}}Unstratified, unmatched WR\end{tabular}    & \begin{tabular}[c]{@{}l@{}}0.04\end{tabular} & \begin{tabular}[c]{@{}l@{}} 0.05\end{tabular} & \begin{tabular}[c]{@{}l@{}}0.05\end{tabular} \\
\begin{tabular}[c]{@{}l@{}}O'Brien's rank-sum-type test\end{tabular}    & \begin{tabular}[c]{@{}l@{}}0.05\end{tabular} & \begin{tabular}[c]{@{}l@{}} 0.05\end{tabular} & \begin{tabular}[c]{@{}l@{}}0.05\end{tabular} \\
\bottomrule
\end{tabular}
\end{adjustbox}
\label{TypeI2}
\end{table}

\textbf{Power for the Same Effects on Both Components} Next, we examine the performance of WR methods by comparing it with other commonly used analyses for cases with either both two components have a similar effect or only one having an effect. Our results are shown in the following tables.

\begin{table}
\caption{The estimation of treatment effect of different sample sizes}
\begin{adjustbox}{width=0.9\columnwidth,center}
\begin{tabular}{@{}llll@{}}
\toprule
Estimation (confidence interval)                                                     & N=60                                                               & N=100                                                              & N=200                                                                                                                       \\ \midrule
\begin{tabular}[c]{@{}l@{}}HR\end{tabular}& \begin{tabular}[c]{@{}l@{}}0.62 (0.36, 1.10)\end{tabular}  & \begin{tabular}[c]{@{}l@{}}0.61 (0.40, 0.93)\end{tabular} & \begin{tabular}[c]{@{}l@{}}0.60 (0.45, 0.82)\end{tabular}  \\
\begin{tabular}[c]{@{}l@{}}stratified, matched WR\end{tabular}& \begin{tabular}[c]{@{}l@{}}1.51 (0.69, 3.87)\end{tabular}    & \begin{tabular}[c]{@{}l@{}}1.49 (0.82, 2.95)\end{tabular} & \begin{tabular}[c]{@{}l@{}}1.49 (0.98, 2.34)\end{tabular}  \\  
\begin{tabular}[c]{@{}l@{}}stratified, unmatched WR\end{tabular} & \begin{tabular}[c]{@{}l@{}}1.59 (0.81, 3.12)\end{tabular}    & \begin{tabular}[c]{@{}l@{}}1.55 (0.95, 2.54)\end{tabular} & \begin{tabular}[c]{@{}l@{}}1.52 (1.08, 2.14)\end{tabular} \\ 
\begin{tabular}[c]{@{}l@{}}unstratified, unmatched WR\end{tabular} & \begin{tabular}[c]{@{}l@{}} 1.55 (0.84, 2.89)\end{tabular}     & \begin{tabular}[c]{@{}l@{}}1.51 (0.94, 2.43)\end{tabular} & \begin{tabular}[c]{@{}l@{}} 1.49 (1.07, 2.08)\end{tabular}\\ 
\bottomrule
\end{tabular}
\end{adjustbox}
\label{SameEffectEstimation}
\end{table}

\begin{table}
\caption{Power comparison with setting $\beta_t=log(0.6)$ and let $\beta_{in}=0$ ($\beta^{\prime}=\beta_t+\beta_{in}=log(0.6)$) to make HR$=0.6$}
\begin{adjustbox}{width=0.6\columnwidth,center}
\begin{tabular}{@{}llll@{}}
\toprule
Power                                                         &N=60                                                               &N=100                                                              & N=200                                                                                                                   \\ \midrule
\begin{tabular}[c]{@{}l@{}}Cox regression\end{tabular}& \begin{tabular}[c]{@{}l@{}}0.44\end{tabular}  & \begin{tabular}[c]{@{}l@{}} 0.66\end{tabular} & \begin{tabular}[c]{@{}l@{}} 0.92\end{tabular}  \\
\begin{tabular}[c]{@{}l@{}}Stratified, matched WR\end{tabular}& \begin{tabular}[c]{@{}l@{}} 0.17\end{tabular}     & \begin{tabular}[c]{@{}l@{}} 0.26\end{tabular} & \begin{tabular}[c]{@{}l@{}}0.47\end{tabular} \\  
\begin{tabular}[c]{@{}l@{}}Stratified,unmatched WR\end{tabular}& \begin{tabular}[c]{@{}l@{}}0.19\end{tabular}     & \begin{tabular}[c]{@{}l@{}}0.36\end{tabular} & \begin{tabular}[c]{@{}l@{}} 0.65\end{tabular} \\
\begin{tabular}[c]{@{}l@{}}Unstratified,unmatched WR\end{tabular}& \begin{tabular}[c]{@{}l@{}}0.21\end{tabular}     & \begin{tabular}[c]{@{}l@{}}0.35\end{tabular} & \begin{tabular}[c]{@{}l@{}} 0.61\end{tabular} \\
\begin{tabular}[c]{@{}l@{}}O'Brien's rank-sum-type test\end{tabular} & \begin{tabular}[c]{@{}l@{}}0.32\end{tabular}     & \begin{tabular}[c]{@{}l@{}}0.51\end{tabular} & \begin{tabular}[c]{@{}l@{}} 0.82\end{tabular}\\
\bottomrule
\end{tabular}
\end{adjustbox}
\label{SameEffectPower}
\end{table}
As seen in Table \ref{SameEffectPower}, it can be observed that the powers order is Cox regression $>$ O'Brien's $>$ stratified unmatched $\sim$ unstratified unmatched $>$ stratified matched when assuming the same effects on both components.

\textbf{Power for Effect on Death Only (no Effect on Hospitalization)}
\begin{table}
\caption{The estimation of treatment effect of different sample sizes}
\centering
\begin{adjustbox}{width=0.9\columnwidth}
\begin{tabular}{@{}llll@{}}
\toprule
Estimation (confidence interval)                                                     & N=60                                                               & N=100                                                              & N=200                                                                                                                       \\ \midrule
\begin{tabular}[c]{@{}l@{}}HR\end{tabular}  & \begin{tabular}[c]{@{}l@{}}0.61 (0.35, 1.09)\end{tabular} & \begin{tabular}[c]{@{}l@{}}0.59 (0.39, 0.91)\end{tabular} & \begin{tabular}[c]{@{}l@{}}0.60 (0.45, 0.81)\end{tabular} \\
\begin{tabular}[c]{@{}l@{}}Stratified, matched WR\end{tabular}    & \begin{tabular}[c]{@{}l@{}}3.02 (1.38, 11.8)\end{tabular} & \begin{tabular}[c]{@{}l@{}}3.06 (1.66, 7.58)\end{tabular} & \begin{tabular}[c]{@{}l@{}}2.98 (1.92, 5.20)\end{tabular} \\  
\begin{tabular}[c]{@{}l@{}}Stratified, unmatched WR\end{tabular}    & \begin{tabular}[c]{@{}l@{}}3.29 (1.58, 6.84)\end{tabular} & \begin{tabular}[c]{@{}l@{}}3.24 (1.87, 5.59)\end{tabular} & \begin{tabular}[c]{@{}l@{}}3.05 (2.10, 4.43)\end{tabular} \\ 
\begin{tabular}[c]{@{}l@{}}Unstratified, unmatched WR\end{tabular}    & \begin{tabular}[c]{@{}l@{}}3.14 (1.59, 6.23)\end{tabular} & \begin{tabular}[c]{@{}l@{}} 3.09 (1.84, 5.21)\end{tabular} & \begin{tabular}[c]{@{}l@{}} 2.96 (2.06, 4.25)\end{tabular} \\ 
\bottomrule
\end{tabular}
\end{adjustbox}
\label{DeathOnlyEstimaiton}
\end{table}

\begin{table}
\caption{Power comparison with setting $\beta_t=0$ and let $\beta_{in}=log(0.18)$ ($\beta^{\prime}=\beta_t+\beta_{in}=log(0.18)$) to make HR$=0.6$ under Cox regression}
\begin{adjustbox}{width=0.6\columnwidth,center}
\begin{tabular}{@{}llll@{}}
\toprule
Power                                                         &N=60                                                               &N=100                                                              & N=200                                                                                                                   \\ \midrule
\begin{tabular}[c]{@{}l@{}}Cox regression\end{tabular}  & \begin{tabular}[c]{@{}l@{}} 0.51\end{tabular} & \begin{tabular}[c]{@{}l@{}} 0.65\end{tabular} & \begin{tabular}[c]{@{}l@{}}0.81\end{tabular} \\
\begin{tabular}[c]{@{}l@{}}Stratified, matched WR\end{tabular}    & \begin{tabular}[c]{@{}l@{}} 0.78\end{tabular} & \begin{tabular}[c]{@{}l@{}}0.94\end{tabular} & \begin{tabular}[c]{@{}l@{}} 0.99\end{tabular} \\  
\begin{tabular}[c]{@{}l@{}}Stratified, unmatched WR\end{tabular}    & \begin{tabular}[c]{@{}l@{}}0.90\end{tabular} & \begin{tabular}[c]{@{}l@{}} 0.99\end{tabular} & \begin{tabular}[c]{@{}l@{}}1\end{tabular} \\
\begin{tabular}[c]{@{}l@{}}Unstratified, unmatched WR\end{tabular}    & \begin{tabular}[c]{@{}l@{}}0.89\end{tabular} & \begin{tabular}[c]{@{}l@{}} 0.99\end{tabular} & \begin{tabular}[c]{@{}l@{}}1\end{tabular} \\
\begin{tabular}[c]{@{}l@{}}O'Brien's rank-sum-type test\end{tabular}    & \begin{tabular}[c]{@{}l@{}}0.50\end{tabular} & \begin{tabular}[c]{@{}l@{}} 0.74\end{tabular} & \begin{tabular}[c]{@{}l@{}}0.93\end{tabular} \\
\bottomrule
\end{tabular}
\end{adjustbox}
\label{DeathOnlyPower}
\end{table}
As seen in Table \ref{DeathOnlyPower}, it can be observed that the powers order is stratified unmatched $>$ unstratified unmatched $\sim$ stratified matched  $>$ O'Brien's $>$ Cox regression.

\textbf{Power for Having Effect on Death Only, but Assuming Wrong Winning Criteria}
\begin{table}
\caption{The estimation of treatment effect of different sample sizes}
\begin{adjustbox}{width=0.9\columnwidth,center}
\begin{tabular}{@{}llll@{}}
\toprule
Estimation (confidence interval)                                                     & N=60                                                               & N=100                                                              & N=200                                                                                                                       \\ \midrule
\begin{tabular}[c]{@{}l@{}}HR\end{tabular}  & \begin{tabular}[c]{@{}l@{}}0.60 (0.34, 1.06)\end{tabular} & \begin{tabular}[c]{@{}l@{}}0.59 (0.39, 0.91)\end{tabular} & \begin{tabular}[c]{@{}l@{}}0.60 (0.44, 0.81)\end{tabular} \\
\begin{tabular}[c]{@{}l@{}}Stratified, matched WR\end{tabular}    & \begin{tabular}[c]{@{}l@{}}1.12 (0.49, 2.65)\end{tabular} & \begin{tabular}[c]{@{}l@{}}1.17 (0.63, 2.22)\end{tabular} & \begin{tabular}[c]{@{}l@{}}1.12 (0.74, 1.73)\end{tabular} \\  
\begin{tabular}[c]{@{}l@{}}Stratified, unmatched WR\end{tabular}    & \begin{tabular}[c]{@{}l@{}}1.19 (0.62, 2.29)\end{tabular} & \begin{tabular}[c]{@{}l@{}}1.19 (0.73, 1.96)\end{tabular} & \begin{tabular}[c]{@{}l@{}}1.15 (0.82, 1.61)\end{tabular} \\ 
\begin{tabular}[c]{@{}l@{}}Unstratified, unmatched WR\end{tabular}    & \begin{tabular}[c]{@{}l@{}}1.18 (0.64, 2.19)\end{tabular} & \begin{tabular}[c]{@{}l@{}} 1.17 (0.73, 1.88)\end{tabular} & \begin{tabular}[c]{@{}l@{}} 1.14 (0.82, 1.58)\end{tabular} \\ 
\bottomrule
\end{tabular}
\end{adjustbox}
\label{WrongCriEstimaiton}
\end{table}

\begin{table}
\caption{Power comparison with setting 
$\beta_t=0$ and let $\beta_{in}=log(0.18)$ ($\beta^{\prime}=\beta_t+\beta_{in}=log(0.18)$) to make HR$=0.6$ under Cox regression}
\begin{adjustbox}{width=0.6\columnwidth,center}
\begin{tabular}{@{}llll@{}}
\toprule
Power                                                         &N=60                                                               &N=100                                                              & N=200                                                                                                                   \\ \midrule
\begin{tabular}[c]{@{}l@{}}Cox regression\end{tabular}  & \begin{tabular}[c]{@{}l@{}} 0.50\end{tabular} & \begin{tabular}[c]{@{}l@{}} 0.66\end{tabular} & \begin{tabular}[c]{@{}l@{}}0.82\end{tabular} \\
\begin{tabular}[c]{@{}l@{}}Stratified, matched WR\end{tabular}    & \begin{tabular}[c]{@{}l@{}} 0.07\end{tabular} & \begin{tabular}[c]{@{}l@{}}0.07\end{tabular} & \begin{tabular}[c]{@{}l@{}} 0.10\end{tabular} \\  
\begin{tabular}[c]{@{}l@{}}Stratified, unmatched WR\end{tabular}    & \begin{tabular}[c]{@{}l@{}}0.06\end{tabular} & \begin{tabular}[c]{@{}l@{}} 0.09\end{tabular} & \begin{tabular}[c]{@{}l@{}}0.12\end{tabular} \\
\begin{tabular}[c]{@{}l@{}}Unstratified, unmatched WR\end{tabular}    & \begin{tabular}[c]{@{}l@{}}0.09\end{tabular} & \begin{tabular}[c]{@{}l@{}} 0.07\end{tabular} & \begin{tabular}[c]{@{}l@{}}0.11\end{tabular} \\
\begin{tabular}[c]{@{}l@{}}O'Brien's rank-sum-type test\end{tabular}    & \begin{tabular}[c]{@{}l@{}}0.51\end{tabular} & \begin{tabular}[c]{@{}l@{}} 0.72\end{tabular} & \begin{tabular}[c]{@{}l@{}}0.91\end{tabular} \\
\bottomrule
\end{tabular}
\end{adjustbox}
\label{WrongCriPower}
\end{table}
As seen in Table \ref{WrongCriPower}, it can be observed that the powers order is  O'brien's $>$ Cox regression $>$ stratified unmatched $\sim$ unstratified unmatched $\sim$ stratified matched when assuming that only effect exists on the death event, not the hospitalization event.

\subsection{Continuous Composite Endpoint With Three Components and Repeated Measurements Under SED}

As highlighted in the introduction, two-stage enrichment designs such as sequential parallel comparison design, SED and sequential multiple assignment randomized trial have been proposed and used in clinical trials. After learning that the use of WR can increase the study power, we are interested in assessing whether the idea of WR can be implemented in two-stage design to further increase trial efficiency for rare disease clinical trials. We consider the SED and compare it with complete randomization (CR) in our evaluation in the followings.

\textbf{Check Type I Error} Drug and placebo are equally effective in all the three components.
\begin{table}
\centering
\caption{Type I error comparison with setting $(p_1,p_2,p_3,p_4)=(0.05,0.05,0.8,0.1)$, $\epsilon\sim N(0,1)$, $\beta_{pj}=\beta_{t1}=-1.5$, $\beta_{in2}=\beta_{in3}=0$, $\beta_{cov1}=\beta_{cov2}=5$, $c_{t}=0.8$, $c_{s0}=0.8$, $c_{s1}=0.9$.}
\begin{adjustbox}{width=0.7\columnwidth}
\begin{tabular}{@{}lllllll@{}}
\toprule
Type I error                 & \multicolumn{2}{c}{N=100}                        & \multicolumn{2}{c}{N=200}               & \multicolumn{2}{c}{N=500}         \\ \midrule
Design                    & \multicolumn{1}{c}{CR} & \multicolumn{1}{c}{SED} & \multicolumn{1}{c}{CR} & \multicolumn{1}{c}{SED} & \multicolumn{1}{c}{CR} & \multicolumn{1}{c}{SED} \\
Coningency table          & 0.05   & 0.05   &0.05  & 0.05          &0.05 &  0.05      \\
Stratified matched WR     & 0.08   & 0.13   &0.07   & 0.07          &0.06 &   0.06     \\
Stratified unmatched WR   & 0.05   & 0.05   &0.06   & 0.04          &0.05 &   0.05     \\
Unstratified unmatched WR & 0.05   & 0.05   &0.06   & 0.04          &0.05 &   0.05    \\
\bottomrule
\end{tabular}
\end{adjustbox}
\label{TypeI_SED}
\end{table}
All Type I errors in \ref{TypeI_SED} are preserved when sample size $N$ is big. In addition, The Type I error under stratified matched WR is preserved more slowly than others.

\textbf{Power Comparison }

\textbf{Scenario 1} The drug is equally effective in improving all three components, and it's more effective than placebo in all the three components. The results are in Table \ref{S1Power_SED}.
\begin{table}
\centering
\caption{Power comparison with setting $(p_1,p_2,p_3,p_4)=(0.05,0.05,0.8,0.1)$, $\epsilon\sim N(0,1)$, $\beta_{pj}=-1.5$, $\beta_{t1}=-2$, $\beta_{in2}=\beta_{in3}=0$, $\beta_{cov1}=\beta_{cov2}=5$, $c_{t}=0.8$, $c_{s0}=0.8$, $c_{s1}=0.9$}
\begin{adjustbox}{width=0.7\columnwidth}
\begin{tabular}{@{}lllllll@{}}
\toprule
Power                 & \multicolumn{2}{c}{N=100}                        & \multicolumn{2}{c}{N=200}               & \multicolumn{2}{c}{N=500}         \\ \midrule
Design                    & \multicolumn{1}{c}{SED} & \multicolumn{1}{c}{CR} & \multicolumn{1}{c}{SED} & \multicolumn{1}{c}{CR} & \multicolumn{1}{c}{SED} & \multicolumn{1}{c}{CR} \\
Coningency table          &0.30               &0.30     &0.58       &0.45   &0.92 &0.90      \\
Stratified matched WR     &0.48               &0.46     &0.77       &0.69   &0.99 &0.99        \\
Stratified unmatched WR   &0.49               &0.47     &0.81       &0.74    &0.99 &0.99        \\
Unstratified unmatched WR &0.33               &0.32     &0.59       &0.51    &0.92 & 0.93       \\
\bottomrule
\end{tabular}
\end{adjustbox}
\label{S1Power_SED}
\end{table}
When $\sum_{j=1}^{3}|\beta_{pj}-\beta_{tj}|=1.5$, SED always outperforms CR. The WR methods for composite components under both designs achieve higher power than other tests when sample size $N$ is large. Stratified methods have higher power than nonstratified methods.

\textbf{Scenario 2} The drug is much more effective than placebo in the first component, but it's equally effective as placebo in the $2^{nd}$ and $3^{rd}$ components. We decrease the drug's overall efficacy to the three components. The results are in Table \ref{S2Power_SED}.
\begin{table}
\centering
\caption{Power comparison with setting $(p_1,p_2,p_3,p_4)=(0.05,0.05,0.8,0.1)$, $\varepsilon\sim N(0,1)$, $\beta_{pj}=-1.5$, $\beta_{t1}=-2$, $\beta_{in2}=\beta_{in3}=0.5$, $\beta_{cov1}=\beta_{cov2}=5$, $c_{t}=0.8$, $c_{s0}=0.8$, $c_{s1}=0.9$.}
\begin{adjustbox}{width=0.7\columnwidth,center}
\begin{tabular}{@{}lllllll@{}}
\toprule
Power                 & \multicolumn{2}{c}{N=100}                        & \multicolumn{2}{c}{N=200}               & \multicolumn{2}{c}{N=500}         \\ \midrule
Design                    & \multicolumn{1}{c}{SED} & \multicolumn{1}{c}{CR} & \multicolumn{1}{c}{SED} & \multicolumn{1}{c}{CR} & \multicolumn{1}{c}{SED} & \multicolumn{1}{c}{CR} \\
Coningency table          & 0.09              & 0.07            &0.16                &0.13           &0.27   &0.20      \\
Stratified matched WR     & 0.15              & 0.14            &0.23                &0.20           &0.40   &0.31        \\
Stratified unmatched WR   & 0.23              & 0.11            &0.27                &0.17           &0.41   &0.32        \\
Unstratified unmatched WR & 0.22              & 0.07            &0.24                &0.14           &0.32   &0.22        \\
\bottomrule
\end{tabular}
\end{adjustbox}
\label{S2Power_SED}
\end{table}
When $\sum_{j=1}^{3}|\beta_{pj}-\beta_{tj}|=0.5$, although powers decrease,  SED still outperforms CR.

\textbf{Scenario 3} We keep assuming that a drug is equally effective in improving the three components and more effective than placebo. However, we adjust patients' distribution, e.g. we decrease the target patient proportion $p_3$. The results are in Table \ref{S3Power_SED}.
\begin{table}
\caption{Power comparison with setting $(p_1,p_2,p_3,p_4)=(0.6,0.05,0.3,0.05)$, $\epsilon\sim N(0,1)$, $\beta_{pj}=-1.5$, $\beta_{t1}=-2$, $\beta_{in2}=\beta_{in3}=0$, $\beta_{cov1}=\beta_{cov2}=5$, $c_{t}=0.8$, $c_{s0}=0.8$, $c_{s1}=0.9$.}
\begin{adjustbox}{width=0.7\columnwidth,center}
\begin{tabular}{@{}lllllll@{}}
\toprule
Power                 & \multicolumn{2}{c}{N=100}                        & \multicolumn{2}{c}{N=200}               & \multicolumn{2}{c}{N=500}         \\ \midrule
Design                    & \multicolumn{1}{c}{SED} & \multicolumn{1}{c}{CR} & \multicolumn{1}{c}{SED} & \multicolumn{1}{c}{CR} & \multicolumn{1}{c}{SED} & \multicolumn{1}{c}{CR} \\
Coningency table          &0.07         &0.06      &0.10     &0.10    &0.20 &0.17     \\
Stratified matched WR     &0.12         &0.07      &0.13     &0.11    &0.23 &0.23        \\
Stratified unmatched WR   &0.23         &0.07      &0.25     &0.15    &0.33 &0.26        \\
Unstratified unmatched WR &0.20         &0.06      &0.24     &0.10    &0.29 &0.19        \\ \bottomrule
\end{tabular}
\end{adjustbox}
\label{S3Power_SED}
\end{table}
Through $\sum_{j=1}^{3}|\beta_{pj}-\beta_{tj}|=1.5$, when sample size $N$ is small, SED even more greatly outperforms CR than the scenario when $p_3=0.8$ .

\section{Conclusion}
The size of clinical trials can be small because of novel or rare diseases or pediatric patient populations. To demonstrate the drug’s effect with substantial evidence, innovative designs, such as two-stage enrichment designs, can be used to enhance trial efficacy. On top of that, using a composite endpoint by incorporating information from different domains of the diseases can be considered. However, how to construct components and combine the information for the composite endpoint is crucial for trial interpretability. The WR methods can be used to further increase study power in a composite endpoint, but correctly specifying a winning criterion is a key to trial success.

In this project, we first examined four types of WR methods, which consider matched or un-matched, and stratified or unstratified for three different types of endpoints (binary, continuous and survival), and have these methods compared with other commonly used tests and approaches; in particular O’Brien's rank-sum type test and Cox regression model for survival endpoints, and contingency table for continuous endpoint. To further enhance trial efficiency, we demonstrated the use of WR method in the SED. A similar analogy can be applied to other types of two staged designs or enrichment designs.

In summary, we found that the stratified unmatched WR method always performs better than the stratified matched WR method. When there is no prior information about winning criteria is available, O’Brien's rank method has greater power than WR methods. Furthermore, using WR methods in two-stage enrichment designs can further enhance trial efficiency.

\bibliographystyle{unsrt}
\bibliography{SBRpaper}

\end{document}